\documentclass{article}%
\usepackage{amsmath}
\usepackage{amsfonts}
\usepackage{amssymb}
\usepackage{graphicx}
\usepackage{booktabs}
\usepackage{array}
\usepackage{rotating}
\usepackage{amsmath}%
\usepackage{ragged2e}
\providecommand{\U}[1]{\protect\rule{.1in}{.1in}}

\begin{document}

\begin{center}
\bigskip 
{\Large
\textbf{Probing Lorentz Invariance Violation in Z Boson Mass Measurements at High-Energy Colliders
}}
\bigskip

\bigskip

{\Huge \bigskip}

\textbf{J. Jejelava}\footnote{juansher.jejelava@iliauni.edu.ge, juansher.jejelava@cern.ch}

\textbf{Z. Kepuladze}\footnote{zurab.kepuladze.1@iliauni.edu.ge, zkepuladze@yahoo.com}

\bigskip

\textit{Institute of Theoretical Physics, Ilia State University, 
0179 Tbilisi, Georgia.}

\textit{and} \textit{Andronikashvili Institute of Physics, TSU, 
0186 Tbilisi, Georgia\ }

\bigskip

\bigskip

\textbf{Abstract}
\end{center}
We propose a minimal extension to the Standard Model by introducing a Lorentz Invariance Violation (LIV) term into the Z boson’s dispersion relation, expressed as $p_{\mu}p^{\mu} = M_{Z}^{2} + \delta_{LIV} (p_{\mu}n^{\mu})^{2}$, where $\delta_{LIV}$ defines the violation scale and $n^{\mu}$ is a unit Lorentz vector specifying the direction of the violation.
This perturbative modification alters the Z boson propagator and decay rate, impacting the Drell-Yan process cross-section at high-energy colliders. Observable effects are most pronounced near the resonance region at high rapidities ($|Y | > 4$), potentially shifting the perceived Z boson mass in general LIV cases and inducing additional sidereal-time modulations for spacelike and lightlike LIV, due to Earth’s rotation relative to distant stars. We outline a targeted
search strategy for ATLAS and CMS, achieving sensitivity to LIV signatures
down to  $|\delta_{LIV}| \approx 10^{-8}$ (or even $10^{-9}$ in an optimistic scenario), offering a fresh perspective on historical and future collider data. Our model predicts
that in experiments with higher collision energies (and thus greater rapidities),
the conventionally reconstructed weak boson mass exhibits a systematic shift.
This could be particularly intriguing in the historical context of Tevatron and
LHC weak boson mass measurements, where an initial discrepancy aligned with
our LIV scenario, although current data from both experiments are now fully
consistent.

\newpage

\section{Introduction}

Lorentz Invariance (LI) is intertwined with almost every aspect of particle
physics and cosmology, and for many researchers, the possibility of Lorentz
Invariance Violation (LIV) is considered speculative and highly unlikely.
There is valid motivation for such beliefs; the Special Theory of Relativity
dictates severe constraints on the nature of our space-time. The high-energy
experiments and cosmic ray observations do not offer much evidence in this
regard either. Specific restrictions could be found in the following
publications: \cite{Colleman-Gleshow, Kostelecky-Russell}. 

High-energy experiments, such as those at the LHC, have largely exhausted significant improvements to the Standard Model (SM). Further improvements are likely to be incremental, as any breakthrough would necessitate higher energies that are unavailable with the current generation of colliders. Given the typical timeframe required to construct large accelerators, we do not anticipate a breakthrough in this regard for the next $10-20$ years. That's why the pursuit of exotic physics is becoming increasingly appealing, as it offers the potential to explore beyond the conventional methodologies of standard physics. Unlike standard approaches, which follow established procedures, exotic physics could leverage new, heuristic, search strategies to significantly enhance the probability of uncovering new phenomena. LIV
represents a prime example of such a topic, making it imperative to seize the opportunity to investigate it using the highest-energy experiments available on Earth, such as those at the LHC.

Cosmic ray observations impose stringent constraints on stable particles
traversing interstellar space; however, evidence from cosmic ray experiments
suggests potential violations of Lorentz Invariance (LIV). For example, the
AGASA experiment detected ultra-high-energy cosmic protons \cite{AGASA} that,
in a Lorentz-invariant framework, should be absent due to the
Greisen-Zatsepin-Kuzmin (GZK) cutoff \cite{GZK}. Additionally, observations of
neutrinos from SN 1987A indicate possible inconsistencies with the LI
\cite{astrophysical_LIV_Limits}. Numerous studies explore this possibility,
including a recent analysis \cite{neutrinos_moura}. In contrast, constraints
on unstable particles are less stringent due to their inherent instability.

Dark Matter Experiments such as the DAMA/LIBRA, also report an annual
modulation of their signal, offering LIV as a potential explanation.

The primary strength of the study presented in this paper lies in its
feasibility using current-generation accelerators. This feasibility stems from
significantly relaxed constraints on neutrinos, as detailed in the following
discussion\footnote{\emph{LIV effects are not generally expected to be similar
across different interactions and particles. Even if they are comparable at
some high energies, they are likely to diverge at lower energies due to
different running.}}. Whether this relaxation is a feature of neutrinos and/or weak interactions remains to be seen. Notably, if these relaxed constraints in neutrino physics extend to the broader weak interaction sector, they could yield effects detectable at the threshold of current LHC energies. At the LHC, the Drell-Yan processes are studied, during which the interaction of the partons of protons produces intermediate bosons that then decay into final states. One important aspect of this process is the physics associated with intermediate weak bosons, which becomes dominant due to the resonance effect near the weak boson mass value. Analyzing this region could provide valuable insights into the possibility of LIV in the weak sector. Therefore, we intend to examine specific aspects of Lorentz Invariance Violation (LIV) physics related to the Z-boson, as the signal and analysis for the neutral boson are cleaner and more tractable than those for the charged W bosons.

\section{Phenomenological LIV expansion of Z-boson physics}

{\normalsize If we imagine that LI is broken and there is a distinct direction
in the 4D space-time given by the unit vector $n_{\mu}$, then anything that
interacts with it in any way should experience Lorentz Invariance Violation
(LIV). In this case, there are only so many ways we can minimally modify the
physics of a vector particle in a non-Lorentz invariant manner. }

{\normalsize One obvious way is to introduce an interaction with matter, which
breaks Lorentz invariance (and perhaps gauge and CPT invariance as well)}

{\normalsize
\begin{equation}
L_{LIV}=\delta_{int}(n^{\mu}Z_{\mu})\overline{\psi}n^{\nu}\gamma_{\nu}\psi
\end{equation}
}

{\normalsize This interaction influences all weak processes and modifies the
cross-section of physical processes in this manner:
\begin{equation}
\sigma_{new}=\sigma_{LI}(1+\delta_{int}\cdot f(\Omega))
\end{equation}
where $f(\Omega)$ is only a function of the special orientation of the
process. Such an effect can only be detected by an experiment with very high
sensitivity to special orientation, requiring the Lorentz breaking parameter
to be no smaller than $10^{-4}\div10^{-5}$ to be visible at the current LHC
\cite{CMS, Lunghi}, which might be too large a value for realistic LIV,
considering restrictions from cosmic ray physics. An important feature of such
an effect is that at all energies, it will always be of the same order of
magnitude comparatively \footnote{\emph{It is assumed that $\delta_{int}$
is constant or scales slowly with energy.}}. }

{\normalsize Additionally, the kinematics of the particles involved in this
interaction will be modified through loop diagrams,
transferring LIV to other particles as well. However, these effects are further
suppressed as higher-order processes. This will farther
push away Lorentz breaking parameter from detectable range due the
cosmological restrictions \cite{Colleman-Gleshow, Kostelecky-Russell}. 
We want to highlight that, despite the LIV interaction parameter potentially being larger than the parameters it induces, such as those influencing the dispersion relation or propagator, surprisingly, searching for some of these induced effects experimentally might be more feasible.}

{\normalsize In contrast to the LIV interaction, modification to the dispersion
relation may prove to be more impactful and easier to discover, as the
threshold energy of the reactions is significantly influenced through
kinematics as well as cross-sections in resonance regions at high energies.
Therefore, we decide to look for possible LIV in the kinematic sector,
specifically in the context of the LHC experiment, and hope to observe some
LIV effect in the resonance region of the Z boson. }

{\normalsize LIV modifications in the quark sector are studied in details in the
context of LHC \cite{Lunghi}. For the Z-boson we can invoke some published
extensions of the Standard Model (SM) for possible Lorentz-violating operators in the kinetic
sector \cite{SME,
Chkareuli-Kepuladze}. However, to explain generally, we can only have these three terms with
the LIV vector $n_{\mu}$ and without introducing higher-order derivatives:
\begin{equation}
\Delta L_{LIV}=\frac{\delta_{LIV}}{2}(\partial_{n}Z^{\mu})(\partial_{n}Z_{\mu
})+\frac{\delta_{1LIV}}{2}(\partial_{\mu}Z_{n})(\partial^{\mu}Z_{n}%
)+\delta_{2LIV}(\partial_{\mu}Z^{\mu})(\partial_{n}Z_{n}) \label{GF}%
\end{equation}
where we have denoted
\begin{equation}
\partial_{n}\equiv n_{\mu}\partial^{\mu},\text{ \ \ \ }Z_{n}\equiv n_{\mu
}Z^{\mu} \label{denote1}%
\end{equation}
}

{\normalsize Within a specific scenario or with any additional condition, we
can only connect $\delta_{iLIV}$ parameters with each other. If we require
gauge invariance, then we must set $\delta_{LIV}=\delta_{1LIV}=-\delta_{2LIV}%
$. These terms play different roles. However due to the nature of kinetic operator, the dispersion relation is affected only by the first term and consequently gives a distinct signature in scattering processes. 
Therefore, in this paper, let us focus only on the first term and accept our model to be the SM in unitary gauge with the addition of $\Delta L_{LIV}$\footnote{\emph{we should also clarify that (\ref{LBL}) is the final effective form in the electroweak broken phase and does not affect Higgs sector.}}.
\begin{equation}
\Delta L_{LIV}=\frac{\delta_{LIV}}{2}\left(  \partial_{n}Z^{\mu}\right)
\left(  \partial_{n}Z_{\mu}\right)  \label{LBL}%
\end{equation}
}

{\normalsize So, (\ref{LBL}) will modify the dispersion relation and
propagator of the vector field. The dispersion relation will take the form:}

{\normalsize
\begin{equation}
k_{\mu}k^{\mu}=M_{eff}^{2}=M_{Z}^{2}+\delta_{LIV}\left(  k_{n}\right)  ^{2}
\label{Mass Shell}%
\end{equation}
where $k_{\mu}$ is the four-momentum of the particle and $k_{n}\equiv n_{\mu
}k^{\mu}$ similar to (\ref{denote1}). $M_{Z}$ is the $Z$ boson mass. Here we
have introduced the notion of effective mass, since threshold energy,
resonance, and generally all the kinematics and momentum conservation are now
determined by $M_{eff}$ instead of $M_{Z}$ \cite{Colleman-Gleshow,
Chkareuli-Kepuladze, Chkareuli}. As for the propagator, we get a fairly simple
expression resembling the massive vector field propagator in the unitary
gauge.%
\begin{equation}
\frac{-i}{k_{\lambda}^{2}-M_{eff}^{2}}\left(  g_{\mu\nu}-\frac{k_{\mu}k_{\nu}%
}{M_{eff}^{2}}\right)  \label{pro0}%
\end{equation}
}

\subsection{{LIV Propagator of the Unstable Particle}}

{\normalsize \bigskip We can not accept (\ref{pro0}) as a full propagator for
our vector boson because near the mass shell (\ref{Mass Shell}) it fails. We
must understand the reason and account for it. Usually, the propagator of the
particle is corrected by self-energy graphs. For the scalar particle it looks
as:%
\begin{equation}
D_{corrected}=\frac{i}{k_{\mu}^{2}-m_{0}^{2}}(1+\frac{\Sigma(k)}{k_{\mu}%
^{2}-m_{0}^{2}}+...)=\frac{i}{k_{\mu}^{2}-m_{0}^{2}-\Sigma(k)}%
\end{equation}
where $m_{0}^{2}$ is a bear mass and $\Sigma(k)$ is a loop diagram. Then, one
fixes the mass of the particle on the regularization momentum $p_{r},$
$\overline{m}^{2}=m_{0}^{2}+\Sigma(p_{r})$. However, in the case of a unstable
particle, $\Sigma(k)$ has an imaginary part, which, using Optical Theorem is
calculated as follows:%
\begin{equation}
\operatorname{Im}\Sigma(k)=k_{0}\Gamma(k)
\end{equation}
where $\Gamma(k)$ is a decay width of the scalar particle. Thus, the corrected
propagator takes the form:%
\begin{equation}
D_{corrected}=\frac{i}{k_{\mu}^{2}-\overline{m}^{2}+ik_{0}\Gamma(k)}%
\end{equation}
Various forms for $\overline{m}^{2}$ are used with different success and accuracy, but in our opinion, the approach of \cite{Willenbrock, Greiner-Muller} appears the most coherent with our purpose.
\begin{equation} 
D_{corrected}=\frac{i}{k_{\mu}^{2}-(m-\frac{i}{2}k_{0}\Gamma(k)/m)^{2}}
\label{proS}%
\end{equation}
}

{\normalsize Additionally, we can observe
\begin{equation}
k_{0}\Gamma(k)=m\Gamma_{0}%
\end{equation}
where $\Gamma_{0}$ is decay rate in comoving reference frame. This gives
$D_{corrected}$ textbook form. }

{\normalsize To repeat the same steps for the vector gauge boson without changing the tensor structure of the vector field propagator, we should naturally obtain the following:
\begin{equation}
D_{corrected}=\frac{-iD_{\mu\nu}}{k_{\mu}^{2}-M_{0}^{2}}\left[  g_{\nu\lambda
}+\frac{\Sigma_{\nu\rho}(k)D_{\rho\lambda}(k)}{k_{\mu}^{2}-M_{0}^{2}%
}+...\right]
\end{equation}
and
\begin{equation}
D_{\mu\nu}(k)\Sigma_{\nu\rho}(k)D_{\rho\lambda}(k)\sim D_{\nu\lambda}(k)
\label{cond1}%
\end{equation}
}

{\normalsize This happens in gauge invariant setup, but when it is
broken by LIV, one has to carefully check the outcome. Without
delving deep into renormalization and loop calculations for the SM modified by
the supposed Lorentz violation, from (\ref{pro0}) in analogy to the derivation
of (\ref{proS}), we can conclude that:
\begin{equation}
D_{corrected}=-i\frac{g_{\mu\nu}-(1+\overline{\delta}_{LIV})k_{\mu}k_{\nu
}/M_{eff}^{2}}{k_{\mu}^{2}-(M_{eff}-ik_{0}\Gamma_{eff}(k)/2M_{eff})^{2}}
\label{pro1}%
\end{equation}
where $\Gamma_{eff}$ is the decay rate of the unstable vector particle and
$\overline{\delta}_{LIV}$ is the modified Lorentz breaking parameter from the
loop calculation. The possible appearance of $\overline{\delta}_{LIV}$ can
only be caused by LIV in the loop diagrams; therefore, it should be at least
suppressed as $\overline{\delta}_{LIV}\sim$ $\alpha\cdot\delta_{LIV}$
($\alpha$ being fine structure constant) and consequently
negligible
\footnote{\emph{Additionally, this }$\overline{\delta}_{LIV}$\emph{
term is proportional to the momentum. Therefore, when }$Z$\emph{-boson interacts with
neutral current, due to its conservation, it may only produce contribution
farther supressed by the ratio }$\dfrac{m_{fermion}^{2}}{M_{Z}^{2}}$,
\emph{which is already way below of the LHC sensitivity}.}.\emph{ } So,
we can safely accept
\begin{equation}
D_{corrected}=-i\frac{g_{\mu\nu}-k_{\mu}k_{\nu}/M_{eff}^{2}}{k_{\mu}%
^{2}-(M_{eff}-ik_{0}\Gamma_{eff}(k)/2M_{eff})^{2}} \label{pro2}%
\end{equation}
}

{\normalsize For the exact expression one has to calculate total vacuum
polarization loop diagrams and find out total effect of the Lorentz violation,
but any possible deviation from (\ref{pro2}), as we mentioned above, has to be
farther suppressed in comparison and inconsequential. }

{\normalsize \bigskip}

\section{Lorentz Violation for Z Boson}

{\normalsize \bigskip In the previous chapter we calculated how propagator
(\ref{pro2}) and dispersion relation (\ref{Mass Shell}) is modified by
the (\ref{LBL}). This will directly affect decay rate (decay width) of the
Weak boson as\ well as corresponding reactions' cross sections. 

In order to calculate the process of our primary interest, the cross-section of the Drell-Yan process via $Z$ boson and neutral current, we need to derive an expression for its decay rate, since it is needed in the (\ref{pro2}).


{\normalsize \bigskip}

\subsection{{\protect\normalsize Modified $Z$ Decay Rate}}

{\normalsize Let us consider the decay of the $Z$ boson into lepton-antilepton
pair to understand how exactly its decay rate is affected by the LIV. If we
denote the $Z$ boson's 4-momentum $k_{\mu}$ and it decays into
lepton-antilepton pair with momentum $p_{\mu}$ and $q_{\mu}$, then the square
of matrix element for the process is
\begin{equation}
\left\vert \mathcal{M}_{el}\right\vert ^{2}=\frac{e^{2}}{4\sin^{2}2\theta_{w}%
}\xi_{\mu}\xi_{\nu}Tr[(\widehat{p}+m_{l})\gamma^{\mu}(g_{l}-\gamma^{5})\left(
\widehat{q}-m_{l}\right)  \gamma^{\nu}(g_{l}-\gamma^{5})]
\end{equation}
}

{\normalsize $\xi_{\mu}$ is a weak boson polarization operator, $e$ is electric charge, $\theta_{w}$ is the Weinberg angle, $m_{l}$ and $g_{l}$ are the mass and coupling to the specific lepton ($m_{l}=0$ and $g_{l}=1$ for neutrino; $g_{l}=1-4\sin^{2} \theta_{w}$ for electron, muon, and tauon). After we sum over polarization of the vector field using (\ref{pro0}) and take into account mass shell conditions and Momentum conservation $k_{\lambda}=p_{\lambda}+q_{\lambda},$ we get following
expression.%
\begin{equation}
\left\vert \mathcal{M}_{el}\right\vert ^{2}=\frac{e^{2}}{\sin^{2}2\theta_{w}%
}\left[  (g_{l}^{2}+1)M_{eff}^{2}+(g_{l}^{2}-2)m^{2}\right]
\end{equation}
where $M_{eff}$ is defined in (\ref{Mass Shell}). }

{\normalsize After account for all averaging and normalization factors, decay
rate takes the form
\begin{align}
\Gamma_{eff}  &  =\frac{1}{96\pi^{2}k_{0}}\int\frac{d^{3}q}{2q_{0}}\frac
{d^{3}p}{2p_{0}}\delta^{4}(k-q-p)\left\vert \mathcal{M}_{el}\right\vert
^{2}\nonumber\\
&  =\frac{e^{2}(g_{l}^{2}+1)}{96\pi^{2}\sin^{2}2\theta_{w}}\frac{1}{k_{0}%
}(M_{eff}^{2}+2\frac{g_{l}^{2}-2}{g_{l}^{2}+1}m^{2})\int d\Omega_{p}%
\frac{p^{2}dp}{q_{0}p_{0}}\delta(k_{0}-q_{0}-p_{0})
\end{align}
}

{\normalsize Phase space integral is Lorentz invariant in the sense that it
can only depend on $k_{\lambda}^{2}$, but $k_{\lambda}^{2}$ is not lorentz
covariant itself on the mass shell (\ref{Mass Shell}). Taking into account
that $p_{0}=\sqrt{\mathbf{p}^{2}+m^{2}}$and $q_{0}=\sqrt{\mathbf{k}%
^{2}+\mathbf{p}^{2}-2\mathbf{kp}\cos\beta+m^{2}},$ ($\beta$ is an angle
between $k_{\mu}$ and $p_{\mu}$ and $\mathbf{k}$, $\mathbf{p}$ are thair
spatial parts), we solve $k_{0}-q_{0}-p_{0}=0$ and get
\begin{align*}
p  &  =\frac{\mathbf{k}M_{eff}^{2}\cos\beta+k_{0}\sqrt{M_{eff}^{4}%
-4m^{2}(k_{0}^{2}-\mathbf{k}^{2}\cos^{2}\beta)}}{2(k_{0}^{2}-\mathbf{k}%
^{2}\cos^{2}\beta)}\\
p_{0}  &  =\frac{k_{0}M_{eff}^{2}+\mathbf{k}\cos\beta\sqrt{M_{eff}^{4}%
-4m^{2}(k_{0}^{2}-\mathbf{k}^{2}\cos^{2}\beta)}}{2(k_{0}^{2}-\mathbf{k}%
^{2}\cos^{2}\beta)}%
\end{align*}
which in the integration over phase space gives
\begin{equation}
\int d\Omega_{p}\frac{p^{2}dp}{q_{0}p_{0}}\delta(k_{0}-q_{0}-p_{0})=\int
d\Omega_{p}\frac{p^{2}}{\left\vert pk_{0}-p_{0}k\cos\beta\right\vert }%
=2\pi\sqrt{1-4\frac{m^{2}}{M_{eff}^{2}}} \label{phase space}%
\end{equation}
}

{\normalsize And finally we get that as expected integral takes "kinda"
Lorentz invariant form. So, we finalize:%
\begin{equation}
\Gamma_{eff}=\frac{\alpha(g_{l}^{2}+1)}{12\sin^{2}2\theta_{w}}\frac{1}{k_{0}%
}(M_{eff}^{2}+2\frac{g_{l}^{2}-2}{g_{l}^{2}+1}m^{2})\sqrt{1-4\frac{m^{2}%
}{M_{eff}^{2}}}%
\end{equation}
where $\alpha$ is the famouse fine-structure constant. If we isolate the first order of LIV effect, we get:}

{\normalsize
\begin{align}
\Gamma_{eff}  &  \approx\Gamma_{SM}\left[  1+\frac{\delta k_{n}{}^{2}}%
{M_{Z}^{2}}\right]  \text{ }=\Gamma_{SM}\frac{M_{eff}^{2}}{M_{Z}^{2}%
}\label{Decay L}\\
\Gamma_{SM}  &  =\frac{\alpha(g_{l}^{2}+1)}{12\sin^{2}2\theta_{w}}\frac
{1}{k_{0}}(M_{Z}^{2}+2\frac{g_{l}^{2}-2}{g_{l}^{2}+1}m^{2})\sqrt
{1-4\frac{m^{2}}{M_{Z}^{2}}}%
\end{align}
where $\Gamma_{SM}$ denotes pure SM value.
In the co-moving frame, during space-like violation $k_{0}=M_{Z}$, even for general violation case, $\Gamma_{0\ eff}\approx\Gamma_{0\ SM}$
\begin{align}
\Gamma_{0\ eff}  & \approx\Gamma_{0\ SM}=\frac{\alpha(g_{l}^{2}+1)}{12\sin^{2}%
2\theta_{w}}\frac{1}{M_{Z}}(M_{Z}^{2}+2\frac{g_{l}^{2}-2}{g_{l}^{2}+1}%
m^{2})\sqrt{1-4\frac{m^{2}}{M_{Z}^{2}}}\nonumber\\
&  \approx\frac{\alpha(g_{l}^{2}+1)}{12\sin^{2}2\theta_{w}}M_{Z}
\label{Decay L0}%
\end{align}
}

{\normalsize Decay rate depends on the scalar product $k_{n}$, and the bigger the momentum is carried by the boson, the bigger is the effect of the LIV. If LIV is time-like, modification is isotropic and decay rate should vary abnormally
with initial energy. In case of space-like LIV, the spatial asymmetry appears in
the decay rate and the modification is maximal, when the initial 3-momentum is
parallel to the preferred direction. While we only explicitly calculated the
lepton channel decay rate, the general form of (\ref{Decay L}) will be
maintained and can be generalized for the total decay rate. }

\subsection{{\protect\normalsize Drell-Yan process}}

{\normalsize \bigskip To adapt effects of (\ref{LBL}) for the experimental testing at LHC, we
need to calculate cross-section of the proton-proton collision, which produces
lepton-antilepton pair through intermediate neutral bosons (Neutral current
Drell-Yan process). According to the parton model, when quark-antiquark
partons are annihilating into some final state $X,$ the cross-section in the
leading order is
}

{\normalsize \begin{flalign*}
\mathrm{\sigma}(pr(P_{1})+pr(P_{2})\rightarrow X)=&&
\end{flalign*}
}

{\normalsize
}

{\normalsize
\[
\sum_{f}\int dx_{1}dx_{2}f_{q_{f}}(x_{1})f_{\overline{q}_{f}}(x_{2}%
)\mathrm{\sigma}(q_{f}(x_{1}P_{1})+\overline{q}_{f}(x_{2}P_{2})\rightarrow X)
\]
where $P_{1,2}$ are initial momentum of the protons, $x_{1,2}$ are portion of
the proton's momentum quark and anti-quark cary with a probability of
$f_{qf}(x_{1})$, $f_{\overline{q}f}(x_{2})$. $f_{qf}(x_{1})$, $f_{\overline
{q}f}(x_{2})$ are parton distribution function-PDF and $f$ denotes the quark
type. }

{\normalsize In our case final state $X$ is lepton-antilepton. 
So, proton-proton scattering cross section, in this channel, is proportional to the cross section of quark-antiquark annihilation. Therefore, we need
$\mathrm{\sigma}(u(x_{1}P_{1})+\overline{u}(x_{2}P_{2})\rightarrow
l+\overline{l}),$ $\mathrm{\sigma}(d(x_{1}P_{1})+\overline{d}(x_{2}%
P_{2})\rightarrow l+\overline{l})$. This process is carried by producing virtual photon, Z-boson and
Higgs after quark-antiquark annihilation, which decays into lepton-antilepton
pair afterward. \ Diagram with intermediate Higgs is extreamely suppressed and
can be safely neglected here. }

{\normalsize At first, lets talk about some kinematic aspects of process.
Momentum caried by intermediate boson we denote by $Q_{\mu}$ and is totaly
determined by the sum of initial momenta.
\begin{equation}
Q=x_{1}P_{1}+x_{2}P_{2}%
\end{equation}
}

{\normalsize When experiment is prepeared in the laboratory frame, we can take
$P_{1,2}$ to be
\begin{equation}
P_{1}=(E,P\overrightarrow{r})\text{, \ \ }P_{2}=(E,-P\overrightarrow{r})
\end{equation}
$\overrightarrow{r}$ is unit vector, which fixes collision axe and is a
direction of the beam (colinear to the detector axe). Since, protons have very
high energies one can neglect proton mass with certain accuracy and assume
$E=P$. So,
\begin{align}
Q_{\mu}  &  =((x_{1}+x_{2})E,(x_{1}-x_{2})P\overrightarrow{r})\\
Q_{\mu}^{2}  &  =4x_{1}x_{2}E^{2}+(x_{1}-x_{2})^{2}m_{p}^{2}%
\end{align}
where $m_{p}$ is proton mass. If\ we neglecte it, then we get familiar form
\begin{align}
Q_{\mu}  &  =E((x_{1}+x_{2}),(x_{1}-x_{2})\overrightarrow{r})\label{Qmiu}\\
Q_{\mu}^{2}  &  =4x_{1}x_{2}E^{2}%
\end{align}
\ In the literature (textbooks), it is accepted to parametrize $Q$ via observebles:
invariant mass $M^{2}$ and rapidity $Y$%
\begin{equation}
Q_{\mu}=M(\cosh Y,\overrightarrow{r}\sinh Y)\text{, \ \ }Q_{\mu}^{2}=M^{2}
\label{Q}%
\end{equation}
where we can note that higher rapidity corresponds with the higher transferred 3-momentum.
}

{\normalsize We can derive the relation for $x_{1}(M, Y)$, $x_{2}(M, Y)$, at energies greater than the mass of proton

\begin{equation}
x_{1,2}=\frac{M}{2E}\exp(\pm Y) \label{x12}%
\end{equation}
and calculate Jacobian
\begin{equation}
\frac{\partial(M^{2},Y)}{\partial(x_{1},x_{2})}=4E^{2}%
\end{equation}
}

{\normalsize This allows us to rewrite
\begin{equation}
\frac{d^{2}\mathrm{\sigma}_{p}}{dM^{2}dY}=\sum_{f}\frac{f_{qf}(x_{1}%
)f_{\overline{q}f}(x_{2})}{4E^{2}}\mathrm{\sigma}_{f}%
\end{equation}
where we have denoted
\begin{align}
\mathrm{\sigma}_{p}  &  =\mathrm{\sigma}(pr(P_{1})+pr(P_{2})\rightarrow
l+\overline{l})\\
\mathrm{\sigma}_{f}  &  =\mathrm{\sigma}(q_{f}(x_{1}P_{1})+\overline{q}%
_{f}(x_{2}P_{2})\rightarrow l+\overline{l})
\end{align}
}

\subsubsection{{\protect\normalsize Cross section of $q+\overline{q}\rightarrow l+\overline{l}$}}

{\normalsize For farther brevity let us denote $\ $%
\[
q_{1,2}^{\mu}=x_{1,2}P_{1,2}^{\mu}%
\]
and if $k_{1,2}^{\mu}=(E_{1,2},\overrightarrow{k}_{1,2})$ are momenta of
lepton-anti lepton pair, we can write that%
\begin{align}
\mathrm{\sigma}_{f}  &  =\frac{1}{4x_{1}x_{2}E^{2}\left\vert \mathrm{v}%
\right\vert (2\pi)^{2}}\int\frac{d^{3}k_{1}d^{3}k_{2}}{4E_{1}E_{2}}\delta
^{4}(Q_{\mu}-k_{1\mu}-k_{2\mu})\left\vert \mathcal{M}_{f}\right\vert ^{2}\\
&  =\frac{1}{(2\pi)^{2}M^{2}\left\vert \mathrm{v}\right\vert }\int\frac
{d^{3}k_{1}}{4E_{1}E_{2}}\delta^{0}(Q_{0}-E_{1}-E_{2})\left\vert
\mathcal{M}_{f}\right\vert ^{2}%
\end{align}
where $\mathit{M}_{f}$ is matrix element for the processand $\left\vert
\mathrm{v}\right\vert $ are relative speed of the partons, which in the first
approximation equals to two. }

{\normalsize 
Since we neglect the mass of the proton for the same reason, we can also neglect the mass of lighter particles.
Therefore, let us also mention that
\begin{equation}
q_{1}^{\mu}q_{2\mu}=q_{1}^{\mu}Q_{\mu}=q_{2}^{\mu}Q_{\mu}=\frac{M^{2}}%
{2},\text{ \ \ }(q_{1,2}^{\mu})^{2}=0
\end{equation}
}

{\normalsize Using (\ref{phase space}), for integration over phase space,
gives
\begin{equation}
\mathrm{\sigma}_{f}=\frac{\left\vert \mathcal{M}_{f}\right\vert ^{2}}{16\pi
M^{2}} \label{cross section}%
\end{equation}
}

{\normalsize For explicit form of cross section we need to calculate matrix
element (including averaging over fermion's helicities).
}

\paragraph{Matrix Element.}
In our analysis of the Drell-Yan process, we adopt the following Z-boson propagator:%
\begin{equation}
D_{\mu\nu}=\frac{i}{Q_{\lambda}^{2}-(M_{eff}-iQ_{0}\Gamma_{eff}(Q_{0}%
)/2M_{eff})^{2}}\left[  g_{\mu\nu}-\frac{Q_{\mu}Q_{\nu}}{M_{eff}^{2}}\right]
\label{pro2a}%
\end{equation}
which can be rewriten using (\ref{Decay L}), (\ref{Decay L0}),
\begin{equation}
\frac{Q_{0}\Gamma_{eff}(k)}{M_{eff}}\approx Q_{0}\Gamma_{SM}(Q_{0}%
)\frac{M_{eff}}{M_{Z}^{2}}\approx\frac{M_{eff}}{M_{Z}}\Gamma_{0SM}%
\end{equation}
even if $Q_{0}\gg M_{Z}$. The $Q_\mu Q_\nu $ term in the expression (\ref{pro2a}) will not contribute in the limit of massless fermions, since
\begin{equation}
Q_{\alpha}J^{\alpha}=Q_{\alpha}J_{A}^{\alpha}=0
\end{equation}
where $J^{\alpha}$ and $J_{A}^{\alpha}$ are vector and axial currents.
Therefore effectively, we can use this propagator
\begin{equation}
D_{\mu\nu}=\frac{ig_{\mu\nu}}{Q_{\lambda}^{2}-M_{eff}^{2}(1-i\Gamma
_{0SM}/2M_{Z})^{2}} \label{pro3}%
\end{equation}

{\normalsize Diagrams involving both the photon and the Z-boson are of the same order of magnitude everywhere except in the resonance region, which is the primary focus of our study, where the Z-boson diagram becomes dominant. By omitting only the Higgs diagram, the resulting matrix element can be expressed as follows:
\[
\mathcal{M}_{f}=\frac{iJ_{e\mu}^{q}J_{e}^{l\mu}}{Q_{\lambda}^{2}}+J_{\mu}%
^{q}J_{\nu}^{l}D^{\mu\nu}%
\]
where $J_{e\mu}^{q}$, $J_{e\mu}^{l}$ are electro-magnetic currents and
\ $J_{\mu}^{q}$ and $J_{\mu}^{l}$ are quark and lepton neutral currents, which
are
\begin{align}
\text{\ \ }J_{\mu}^{l}  &  =\frac{-e}{2\sin2\theta_{w}}\overline{\psi}%
_{l}\gamma_{\mu}(g_{l}-\gamma_{5})\psi_{l}\text{, \ }g_{l}=1-4\sin^{2}%
\theta_{w}\\
J_{\mu}^{u}  &  =\frac{e}{2\sin2\theta_{w}}\overline{u}\gamma_{\mu}%
(g_{u}-\gamma_{5})u\text{, \ }g_{u}=1-\frac{8}{3}\sin^{2}\theta_{w}\\
J_{\mu}^{d}  &  =\frac{-e}{2\sin2\theta_{w}}\overline{d}\gamma_{\mu}%
(g_{d}-\gamma_{5})d\text{, }g_{d}=1-\frac{4}{3}\sin^{2}\theta_{w}%
\end{align}
with $e$ being electric charge and $\theta_{w}$ Weinberg angle. \ }

{\normalsize By introducing Mandelstam variable $S$%
\begin{equation}
S=M^{2}=Q_{\mu}^{2}=(q_{1}+q_{2})_{\mu}^{2}=(p_{1}+p_{2})_{\mu}^{2}%
\end{equation}
the matrix element takes familiar form
\begin{align}
\left\vert \mathcal{M}_{f}\right\vert ^{2}  &  =\frac{16e^{4}e_{f}^{2}}%
{3}+\frac{8e^{4}\left\vert e_{f}\right\vert g_{q}g_{l}\left(  S-M_{eff}%
^{2}(1-\Gamma_{0SM}^{2}/4M_{Z}^{2})\right)  }{3S\sin^{2}2\theta_{w}%
}R(S)\nonumber\\
&  +\frac{e^{4}(1+g_{q}^{2})(1+g_{l}^{2})}{3\sin^{4}2\theta_{w}}%
R(S)\label{Mel}\\
R(S)  &  =\frac{S^{2}}{\left(  S-M_{eff}^{2}(1-\Gamma_{0SM}^{2}/4M_{Z}%
^{2})\right)  ^{2}+M_{eff}^{4}\Gamma_{0SM}^{2}/M_{Z}^{2}} \label{res factor}%
\end{align}
where $\left\vert e_{f}\right\vert $ equals to $2/3$ or $1/3$ respectively for
$u$ and $d$ quarks. Also, for the brevity, we have introduced resonance factor
$R(S)$. Consequently for the cross-section (\ref{cross section}) we get:%
\begin{align}
\mathrm{\sigma}_{f}  &  =\mathrm{\sigma}_{EM}[1+\frac{g_{q}g_{l}\left(
S-M_{eff}^{2}(1-\Gamma_{0SM}^{2}/4M_{Z}^{2})\right)  }{2\left\vert
e_{f}\right\vert S\sin^{2}2\theta_{w}}R(S)\nonumber\\
&  +\frac{(1+g_{q}^{2})(1+g_{l}^{2})}{16e_{f}^{2}\sin^{4}2\theta_{w}%
}R(S)]\label{CS flavor}\\
\mathrm{\sigma}_{EM}  &  =\frac{4\pi\alpha^{2}}{9S}e_{f}^{2}%
\end{align}
}

{\normalsize Finally, for the differential cross-section for the
proton-proton collision is:%
\begin{align}
\frac{d^{2}\mathrm{\sigma}_{p}}{dSdY}  &  =\sum_{f}\frac{f_{qf}(x_{1}%
)f_{\overline{q}f}(x_{2})}{4E^{2}}\mathrm{\sigma}_{EM}[1+\frac{g_{q}%
g_{l}\left(  S-M_{eff}^{2}(1-\Gamma_{0SM}^{2}/4M_{Z}^{2})\right)
}{2\left\vert e_{f}\right\vert S\sin^{2}2\theta_{w}}R(S)\nonumber\\
&  +\frac{(1+g_{q}^{2})(1+g_{l}^{2})}{16e_{f}^{2}\sin^{4}2\theta_{w}}R(S)]
\label{D cross-section}%
\end{align}
}

{\normalsize Now, it is time to make some conclusions and predictions. In the standard LI setup, this cross-section experiences resonance growth and peaks at the resonance value $S_{r}=M_{Z}^{2}(1-\Gamma_{0SM}^{2}/4M_{Z}^{2})$. The mass and the decay rate define
resonance value and the maximum of the cross section is being defined by the
$1/\Gamma_{0SM}^{2}$. However, during LIV, the resonance mechanism of the differential cross-section (\ref{D cross-section}) is modified, becoming strongly dependent on the transferred momentum. The new resonance maximum is located at:
\begin{equation}
S_{r}=M_{eff}^{2}(1-\Gamma_{0SM}^{2}/4M_{Z}^{2}) \label{Sr0}%
\end{equation}
and the peak value for the cross-section, where we can neglect electromagnetic
part, becomes:}

{\normalsize
\begin{equation}
\mathrm{\sigma}_{f\max}\approx\mathrm{\sigma}_{f\max}^{LI}(1-\frac
{\delta_{LIV}\left(  n\cdot Q_{r}\right)  ^{2}}{M_{Z}^{2}}) \label{R1}%
\end{equation}
where $Q_{r}$ stands for resonance value of the momentum. }

{\normalsize So, basically by measuring $S_{r},$ we measure effective mass,
which is function of energy, in case of $\delta_{LIV}>0$, with higher transferred momentum peak moves right and decreases in height.}

Since the Lorentz-invariance violation (LIV) effect depends on \( n_{\mu} \), three distinct cases can be identified: time-like, space-like, and light-like, defined as follows:
\begin{align}
\text{Time-like:} \quad n_{\mu} &= (1, \vec{0}), \\
\text{Space-like:} \quad n_{\mu} &= (0, \vec{n}), \quad \text{with} \quad \vec{n}^2 = 1, \\
\text{Light-like:} \quad n_{\mu} &= (1, \vec{n}).
\end{align}
These cases are analyzed separately below.

\subsubsection{{\protect\normalsize Time-like violation case}}

{\normalsize For the time-like case, (\ref{Q}) will lead us to
\begin{equation}
n\cdot Q=M\cosh Y=\sqrt{S}\cosh Y
\end{equation}
which gives expresion for the effective mass
\begin{equation}
M_{eff}^{2}=M_{Z}^{2}+\delta_{LIV}\cdot S\cosh^{2}Y
\end{equation}
ready to apply to (\ref{D cross-section}). The expression (\ref{Sr0}) will change accordingly}

{\normalsize
\begin{align}
S_{r}  &  \approx M_{Z}^{2}(1+\delta_{LIV}\cosh^{2}Y)-\Gamma_{0SM}^{2}/4\\
\mathrm{\sigma}_{f\max}  &  \approx\mathrm{\sigma}_{f\max}^{LI}(1-\delta
_{LIV}\cosh^{2}Y)
\end{align}
}

We observe that LIV should be amplified by the rapidity $Y$ of the $Z$-boson, while staying independent of spacial orientation of the experiment. If the data is sorted by the rapidity $Y$, then low rapidity data should allow better determination of the $M_{Z}^{2}$ through resonance value $S_{r}$, while with bigger $Y,$ LIV effect kicks in offering possibility of observing it on the experiment. If one is to constrain $\delta_{LIV}$ from the $Z$ mass uncairtentity $\Delta M_{Z}$ (Atlas value), we can say that
\begin{equation}
\delta_{LIV}\leq\frac{\Delta M_{Z}}{M_{Z}\cosh^{2}Y}%
\end{equation}
which for $Y=5,6$, offers $10^{-8}(10^{-9})$.

We observe that LIV should be amplified by the rapidity $Y$ of the $Z$-boson, while staying independent of spacial orientation of the experiment.
Sorted by the rapidity, data allow better determination of  $M_{Z}^{2}$  through resonance value $S_{r}$ for low values of Y. The LIV effects fully kick in only for high values of rapidity. To constrain $\delta_{LIV}$ withing the precision of the experiment, we can use following relation:
\begin{equation}
\delta_{LIV}\leq\frac{\Delta M_{Z}}{M_{Z}\cosh^{2}Y}%
\end{equation}
which for $Y=5$, offers $10^{-8}(10^{-9})$.

\subsubsection{{\protect\normalsize Space-like violation case}}

{\normalsize For the space-like violation, we have
\begin{equation}
M_{eff}^{2}=M_{Z}^{2}+\delta_{LIV}\cdot S\sinh^{2}Y\cos^{2}\beta
\end{equation}
$\beta$ being an angle between Lorentz violation direction and axe of
collision (i.e. $\overrightarrow{n}$ and $\overrightarrow{r}$). Then for the
resonance we have%
\begin{align}
S_{r}  &  \approx M_{Z}^{2}(1+\delta_{LIV}\sinh^{2}Y\cos^{2}\beta
)-\Gamma_{0SM}^{2}/4\\
\mathrm{\sigma}_{f\max}  &  \approx\mathrm{\sigma}_{f\max}^{LI}(1-\delta
_{LIV}\sinh^{2}Y\cos^{2}\beta)
\end{align}
}

During space-like violation, like the time-like case, dependence on the rapidity $Y$ is strong, but in addition, now we have variation with respect to the orientation of the experiment, and since it rotates with Earth, hourly (sidereal) modulation is expected in the experiment data.

\subsubsection{{\protect\normalsize Light-like violation case}}

{\normalsize For the Light-like violation, we have
\begin{equation}
M_{eff}^{2}=M_{Z}^{2}+\delta_{LIV}\cdot S\left(  \cosh Y-\sinh Y\cos
\beta\right)  ^{2}%
\end{equation}
which is kind of hybrid case, but with a slightly more complicated behavior. }

{\normalsize
\begin{align}
S_{r}  &  \approx M_{Z}^{2}(1+\delta_{LIV}\left(  \cosh Y-\sinh Y\cos
\beta\right)  ^{2})-\Gamma_{0SM}^{2}/4\\
\mathrm{\sigma}_{f\max}  &  \approx\mathrm{\sigma}_{f\max}^{LI}(1-\delta
_{LIV}\left(  \cosh Y-\sinh Y\cos\beta\right)  ^{2})
\end{align}
}

Unlike the space-like case, the LIV effect peaks when $n_\mu$ is perpendicular to the colliding beam, while still exhibiting time modulation\footnote{\textit{In principal, light-like behavior is similar to what general case
scenario would look like. For$\ \ n_{\mu}=(n_{0},\overrightarrow{n})$, with
$\left\vert n_{0}^{2}-\overrightarrow{n}^{2}\right\vert =1$. 
\begin{align*}
M_{eff}^{2}  &  =M_{Z}^{2}(1+\delta_{LIV}\left(  n_{0}\cosh Y-\left\vert
\overrightarrow{n}\right\vert \sinh Y\cos\beta\right)  ^{2})\\
\mathrm{\sigma}_{f\max}  &  \approx\mathrm{\sigma}_{f\max}^{LI}(1-\delta
_{LIV}\left(  n_{0}\cosh Y-\left\vert \overrightarrow{n}\right\vert \sinh
Y\cos\beta\right)  ^{2})
\end{align*}
with strong dependence both on rapidity and special orientation. $n_{0}\gg1$ gives light-like limit and $n_{0}\approx1$ or
$\left\vert \overrightarrow{n}\right\vert \approx1$ time-like and space-like limits.}}}.

{\normalsize \newpage}

\section{Experimental signature of LIV}

As we see from the form of the cross-section (\ref{CS flavor}),
if LIV takes place, it strongly depends on two factors: the rapidity of the
Z-boson and the orientation of the experiment, since the momentum of the
intermediate Z-boson is practically parallel to the axis of the detector. The
LIV effect is strongly pronounced in higher rapidity cases within the
resonance region of the intermediate boson. While different types of
violations share similar general features, space-like and light-like
violations are more challenging to analyze. Therefore, let us focus on the
time-like violation case to highlight the general characteristics of LIV.

Until now, we have not discussed the sign of $\delta_{LIV}$, as it has not
been necessary. However, from (\ref{Mass Shell}, \ref{Sr0}%
, \ref{R1}) we understand that it defines several features differently. We can isolate the effects of violation by plotting the relative differ-
ence between the LIV (\ref{CS flavor}) and the standard (LI) cross-sections, [Fig.\ref{fig:liv_correction_isolated}].

{\normalsize \begin{figure}[ptbh]
\caption{$(\sigma_{LIV}-\sigma_{LI})/\sigma_{LI}$}%
\label{fig:liv_correction_isolated}%
{\normalsize  \centering
\includegraphics[width=0.9\textwidth]{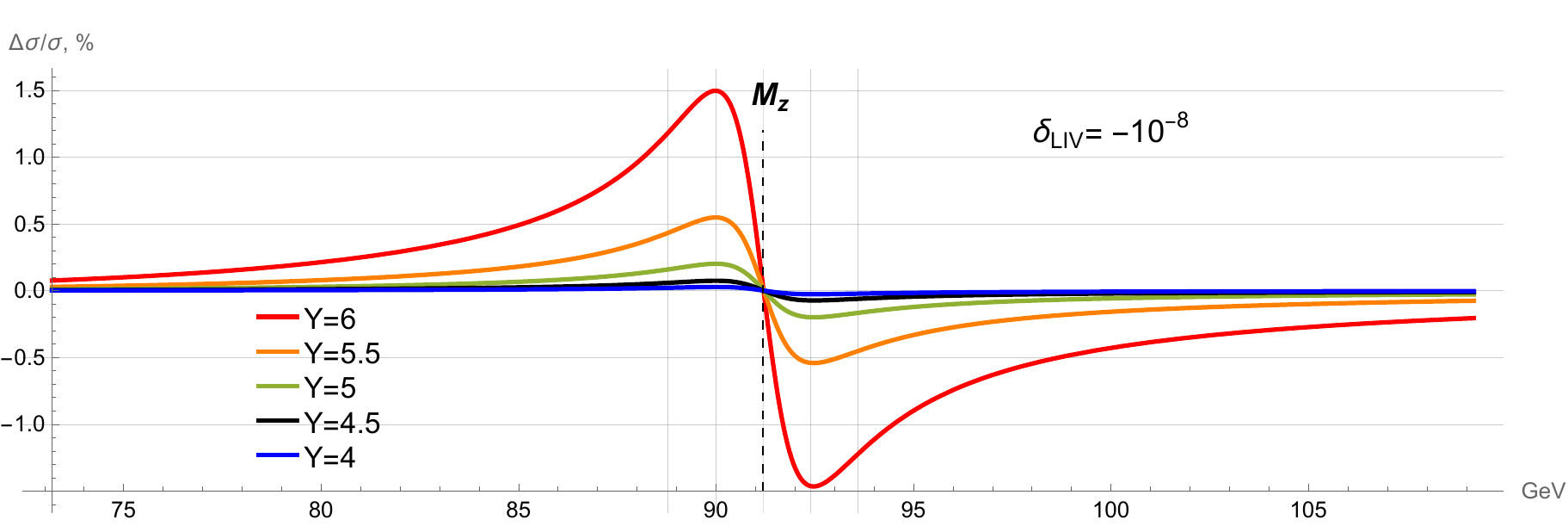}
\includegraphics[width=0.9\textwidth]{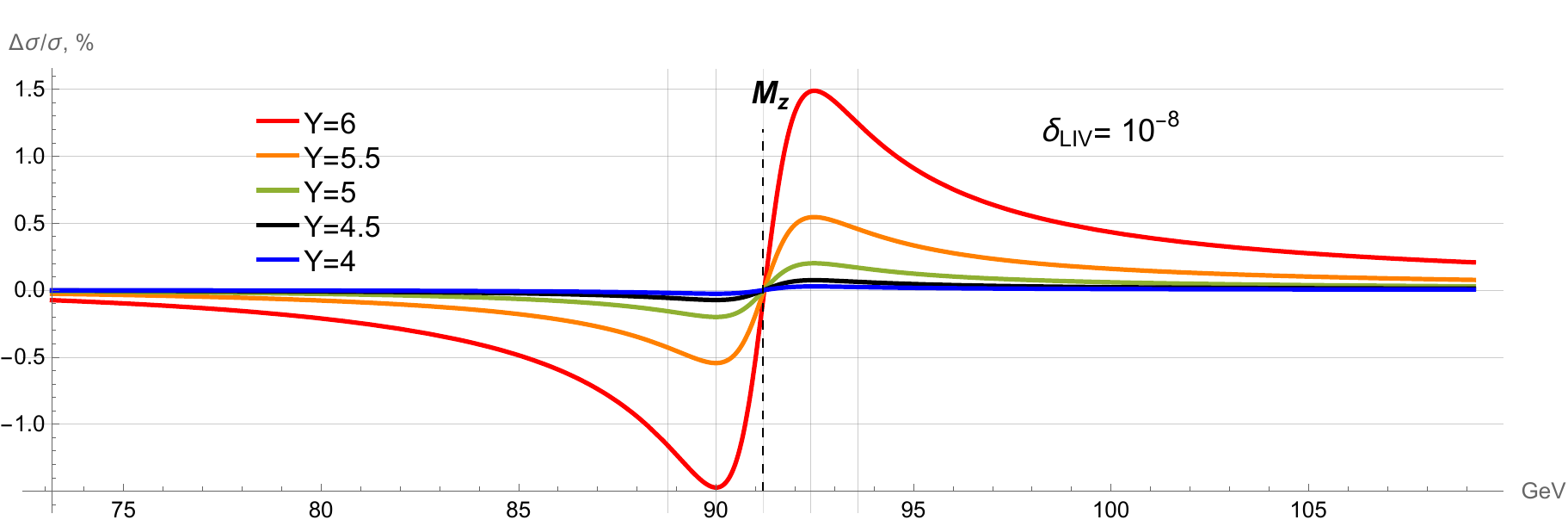}  }
\par
\begin{flushleft}
{\justify {\footnotesize The plot illustrates the difference between $\sigma_{LIV}$ (the Lorentz-violating cross-section) and $\sigma_{LI}$ (the Lorentz-invariant cross-section, equivalent to the Standard Model cross-section).
By dividing this difference by the LI cross-section, we can isolate the LIV effects to better understand their structure. Despite its visual appearance, the LIV effect is not exactly zero at the resonance point, which becomes clear in an exaggerated Figure [Fig. \ref{fig:liv_corrections}]. The LIV effect on the cross-section reaches its maximum value at a point approximately $1.2\  GeV$ away from the true mass ($M_Z = 91.1876\ GeV$). This shift is expected to cause a resonance mass deviation during the standard fitting procedure if it is not adjusted for the distorted shape of the LIV cross-section. The sign of $\delta_{LIV}$ determines whether the shift causes an overestimation or underestimation of the resonance mass relative to $M_Z$. }  }
\end{flushleft}
\end{figure}}

\begin{figure}[ptbh]
\caption{Highly exaggerated comparison of LIV and LI cross-sections.}%
\label{fig:liv_corrections}%
\centering
\includegraphics[width=1\textwidth]{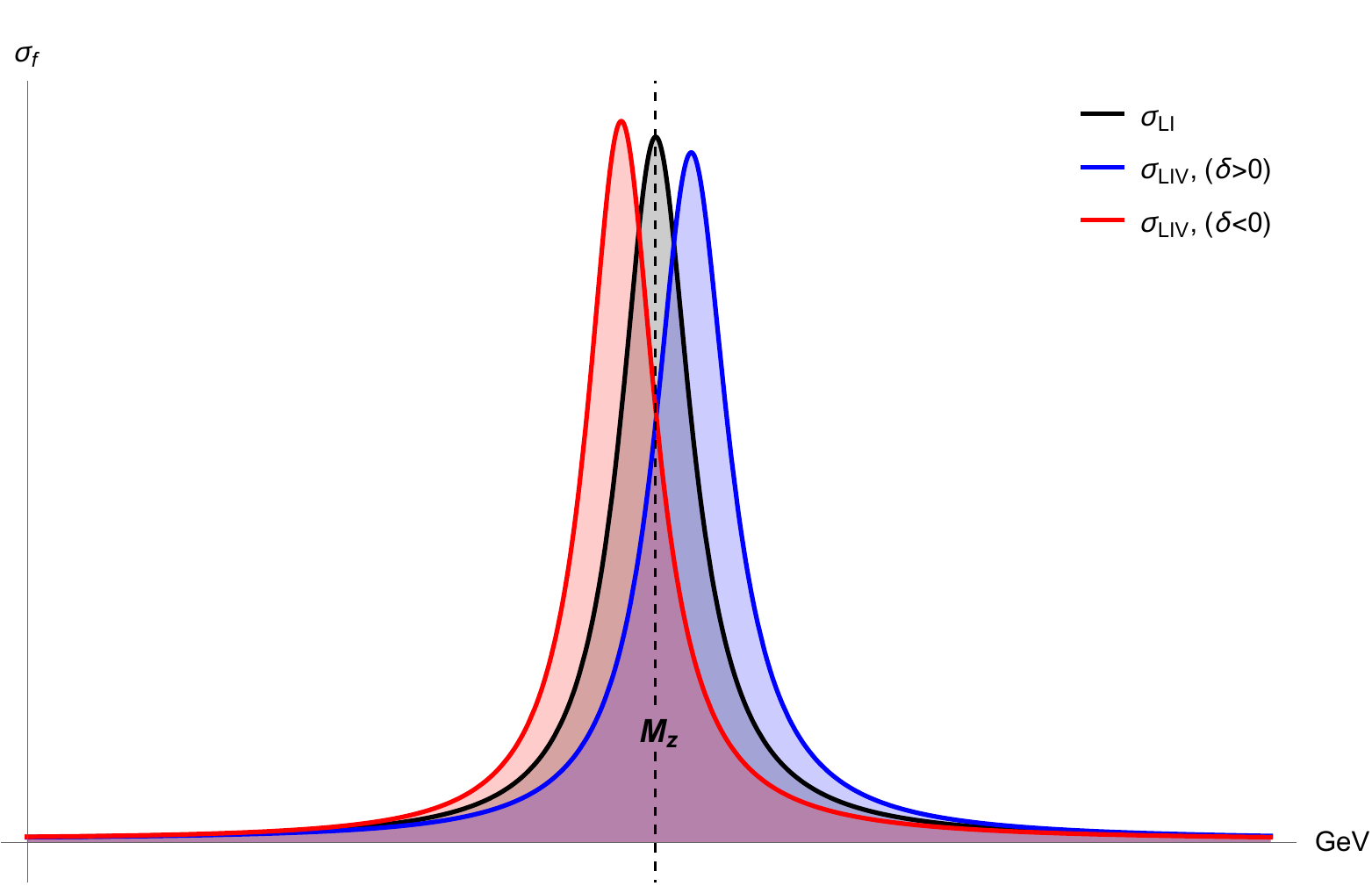}
\includegraphics[width=1\textwidth]{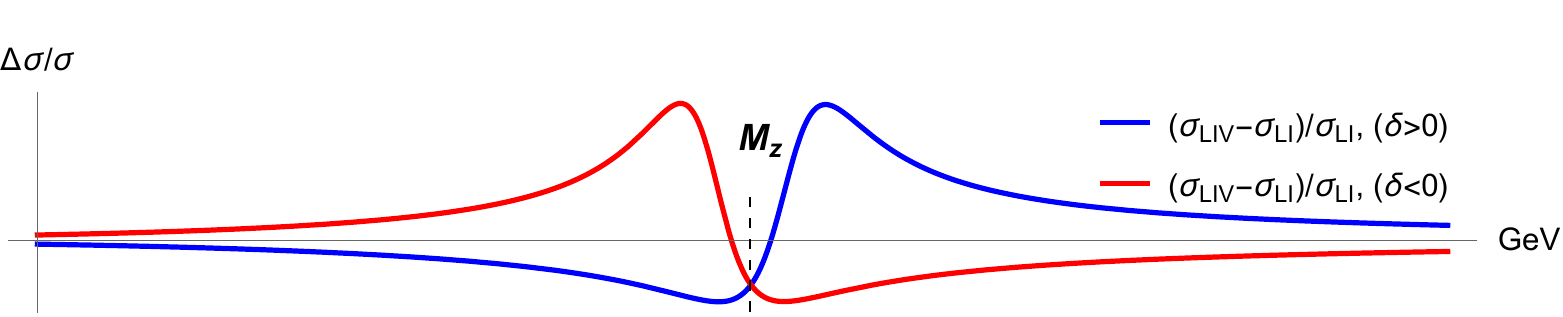}
\par
\begin{flushleft}
{\footnotesize This $Y=8$ plot is for demonstrative purposes only. For the LHC data, the expected difference is limited to $\sim 0.1\%$ for the rapidity $Y=4.5$ , rendering it practically indistinguishable
by visual inspection alone.}
\end{flushleft}
\end{figure}

Figure \ref{fig:liv_correction_isolated} illustrates how the sign of $\delta_{LIV}$ influences the outcome. In both cases, the magnitude of the LIV effect is nearly identical, but its sign differs. LIV effects show maximal deviations from Lorentz invariance in the resonance region, rapidly diminishing farther from it. For $\left|\delta_{LIV}\right|=10^{-8}$, the LIV effect increases by a few tenths of a percent at $Y=5$ and up to 1.5\% at $Y=6$, relevant for future experiment energies. Even small LIV effects, such as 0.2\%, remain significant. Although these differences are too subtle to discern on the paper's scale, for demonstration purposes, we provide an exaggerated plot to highlight the LIV effect's structure in contrast to the Lorentz-invariant cross-section [Fig. \ref{fig:liv_corrections}].

{\normalsize
}

{\normalsize Despite of the fact that LIV becomes significant for high
rapidities, the number of such events are relatively small and, during the typical data collection, all events are gathered together, regardless of LIV effects. Therefore, the majority of events contain an unnoticeable amount of LIV data. Imagine a scenario where an analysis team attempts to find the resonance mass without knowing the LIV nature of the data. The result would be an attempt to fit the affected data to the standard cross-section. Because of the perturbative nature of the Lorentz-violating scenario drastic changes is precluded and the most likely result would be a change in the reconstructed mass and possibly decay rate, since they defined peak and general shape of the
resonance region. To better understand this behavior, we compare $\Delta M_Z$ for different mixtures of LI and LIV data, with results presented in the following table [Tab. \ref{tab:z_mass_shift}]. }

{\normalsize \begin{table}[ptbh]
\caption{Absolute mass shift $|\Delta M_{Z}|$ as a function of rapidity $Y$
and the fractional composition of LI and LIV contributions of cross sections
in the data. \newline}%
\label{tab:z_mass_shift}%
{\normalsize  \centering
\resizebox{\textwidth}{!}{\tiny \setlength{\tabcolsep}{4pt}
\begin{tabular}{>{\centering\arraybackslash}p{1cm}*{11}{>{\raggedleft\arraybackslash}p{1cm}}}
\toprule
${LI}  :$& 100\% & 90\% & 80\% & 70\% & 60\% & 50\% & 40\% & 30\% & 20\% & 10\% & 0\% \\
${LIV} :$& 0\% & 10\% & 20\% & 30\% & 40\% & 50\% & 60\% & 70\% & 80\% & 90\% & 100\% \\
\midrule
Y=0.5 & 0\text{eV} & 0.1\text{keV} & 0.1\text{keV} & 0.2\text{keV} & 0.2\text{keV} & 0.3\text{keV} & 0.3\text{keV} & 0.4\text{keV} & 0.5\text{keV} & 0.5\text{keV} & 0.6\text{keV} \\
Y=1.0 & 0\text{eV} & 0.1\text{keV} & 0.2\text{keV} & 0.3\text{keV} & 0.4\text{keV} & 0.5\text{keV} & 0.7\text{keV} & 0.8\text{keV} & 0.9\text{keV} & 1.0\text{keV} & 1.1\text{keV} \\
Y=1.5 & 0\text{eV} & 0.3\text{keV} & 0.5\text{keV} & 0.8\text{keV} & 1.0\text{keV} & 1.3\text{keV} & 1.5\text{keV} & 1.8\text{keV} & 2.\text{keV} & 2.3\text{keV} & 2.5\text{keV} \\
Y=2.0 & 0\text{eV} & 0.6\text{keV} & 1.3\text{keV} & 1.9\text{keV} & 2.6\text{keV} & 3.2\text{keV} & 3.9\text{keV} & 4.5\text{keV} & 5.2\text{keV} & 5.8\text{keV} & 6.5\text{keV} \\
Y=2.5 & 0\text{eV} & 1.7\text{keV} & 3.4\text{keV} & 5.1\text{keV} & 6.9\text{keV} & 8.6\text{keV} & 10.3\text{keV} & 12.\text{keV} & 13.7\text{keV} & 15.4\text{keV} & 17.1\text{keV} \\
Y=3.0 & 0\text{eV} & 4.6\text{keV} & 9.2\text{keV} & 13.9\text{keV} & 18.5\text{keV} & 23.1\text{keV} & 27.7\text{keV} & 32.4\text{keV} & 37.\text{keV} & 41.6\text{keV} & 46.2\text{keV} \\
Y=3.5 & 0\text{eV} & 12.5\text{keV} & 25.1\text{keV} & 37.6\text{keV} & 50.1\text{keV} & 62.6\text{keV} & 75.2\text{keV} & 87.7\text{keV} & 100.2\text{keV} & 112.7\text{keV} & 125.3\text{keV} \\
Y=4.0 & 0\text{eV} & 34\text{keV} & 68\text{keV} & 102\text{keV} & 136\text{keV} & 170\text{keV} & 204\text{keV} & 238\text{keV} & 272\text{keV} & 306\text{keV} & 340\text{keV} \\
Y=4.5 & 0\text{eV} & 92\text{keV} & 185\text{keV} & 277\text{keV} & 370\text{keV} & 462\text{keV} & 0.6\text{MeV} & 0.6\text{MeV} & 0.7\text{MeV} & 0.8\text{MeV} & 0.9\text{MeV} \\
Y=5.0 & 0\text{eV} & 251\text{keV} & 0.5\text{MeV} & 0.8\text{MeV} & 1.0\text{MeV} & 1.3\text{MeV} & 1.5\text{MeV} & 1.8\text{MeV} & 2.0\text{MeV} & 2.3\text{MeV} & 2.5\text{MeV} \\
Y=5.5 & 0\text{eV} & 0.7\text{MeV} & 1.4\text{MeV} & 2.0\text{MeV} & 2.7\text{MeV} & 3.4\text{MeV} & 4.1\text{MeV} & 4.8\text{MeV} & 5.5\text{MeV} & 6.1\text{MeV} & 6.8\text{MeV} \\
Y=6.0 & 0\text{eV} & 1.9\text{MeV} & 3.7\text{MeV} & 5.6\text{MeV} & 7.4\text{MeV} & 9.3\text{MeV} & 11.1\text{MeV} & 13.0\text{MeV} & 14.8\text{MeV} & 16.7\text{MeV} & 18.6\text{MeV} \\
Y=6.5 & 0\text{eV} & 5\text{MeV} & 10\text{MeV} & 15\text{MeV} & 20\text{MeV} & 25\text{MeV} & 30\text{MeV} & 35\text{MeV} & 40\text{MeV} & 45\text{MeV} & 51\text{MeV} \\
Y=7.0 & 0\text{eV} & 14\text{MeV} & 27\text{MeV} & 41\text{MeV} & 55\text{MeV} & 69\text{MeV} & 82\text{MeV} & 96\text{MeV} & 110\text{MeV} & 124\text{MeV} & 137\text{MeV} \\
\bottomrule
\end{tabular}
}  }
\par
\begin{flushleft}  {\footnotesize {\justify
This table presents the resonance mass shifts obtained when the $\sigma_{LIV}$ cross-section is reconstructed using a Standard Model fit. The notation "LI:" refers to the Standard Model cross-section (without Lorentz violation), while "LIV:" denotes the cross-section affected by LIV effects. The total experimental data is assumed to be a mixture of low-rapidity events (containing undetectable perturbations from the Standard Model) and high-rapidity events (exhibiting detectable LIV effects, manifested as a resonance mass shift from the true mass).
Although real data will consist of a distribution of events across all possible rapidities, for demonstrative purposes, we simplify the data composition to a mixture of LI and LIV data. As expected, the magnitude of the mass shift depends on the level of LIV
contamination within the overall data. The sign of $\delta_{LIV}$ leads to an overestimation ($\delta_{LIV} > 0$) or underestimation ($\delta_{LIV} < 0$) of the resonance mass, but the difference remains within the margin of displayed precision, that’s why for the sake of simplification, we combine both cases into one chart.

\textit{Once again, we emphasize that these resonance mass shifts would occur within a standard analysis framework due to the fitting of an LI cross-section function to an LIV-distorted distribution. Consequently, the LIV contamination affects the fitting parameters of conventional analysis methods.} }}

\end{flushleft}
\par
{\normalsize
}\end{table}}

From the table it is obvious what have talked about already many
times: the low rapidity cases provide no detectable results, while for high
rapidities they may appear. When the data is a mixture of 90\% LI and 10\% LIV events  (for the experimentally feasible $Y=5$), the estimated mass shift is $\left\vert {\normalsize \Delta M_{z}}\right\vert \approx0.25$
$MeV$. However, if one isolates only LIV data (for the same $Y=5$), the mass shift increases to $\left\vert \Delta M_{z}\right\vert \approx 2.5 \ MeV$, too big to be accommodated within experiment uncertainty since declared accuracy is 2.1 MeV, for both Tevatron and LHC. The deviation for the rapidities $Y\geqslant6$ (unfortunately unavailable for the current generation of accelerators) is naturally more pronounced and easier to identify.

Taking into account experimental values for the parton
distribution function would allow for more accurate analysis and expected to be performed in the next publication. Nevertheless, while incorporating parton distribution functions may refine the exact numbers, it does not alter the general nature of our conclusions. In standard data analysis, the limited amount of LIV data typically results in only marginal effects. However, when Y=4,5 cases are analyzed separately, the LIV effect is distinctly evident in the data for $\left|\delta_{LIV}\right| = 10^{-8}$, and possibly for $\left|\delta_{LIV}\right| = 10^{-9}$. If Lorentz invariance violation (LIV) occurs within this range, next-generation accelerators would likely detect it, as the effect becomes highly pronounced at $Y \geqslant 6$. Nevertheless, our analysis remains valuable for current accelerators, as the predicted mass shifts, lying at the edge of measurement precision, open new possibilities for comparing data from accelerators with differing collision energies, such as the Tevatron and LHC.

In the LIV scenario we consider, accelerators with higher collision energies may systematically underestimate the Z boson mass for negative $\delta_{LIV}$ due to varying mixtures of rapidities. We initially chose the Z boson as the focus of our research because of its higher precision in mass measurements. However, we expect a similar effect for the W boson, as our approach should apply to the entire weak vector multiplet. This is particularly intriguing from a historical perspective, as LIV offers a compelling explanation for data discrepancies among the CDF, ATLAS, and CMS experiments. Indeed, for negative $\delta_{LIV}$, higher-energy colliders would experience greater LIV data contamination, leading to a more pronounced systematic underestimation of $M_W$. This aligns with historical data, where, until recently, measurements showed discrepancies: $M_{W\ CDF(2022)} = 80,433.5 \pm 9.4\ MeV$ \cite{MW:CDF}, $M_{W\ CMS(2024)} = 80,360.2 \pm 9.9\ MeV$ \cite{MW:CMS}, and $M_{W\ ATLAS(2023)} = 80,360 \pm 16\ MeV$ \cite{MW:ATLAS}. Recent LHC analyses have shown improved agreement with earlier Tevatron results for the W boson mass. However, this convergence has prompted some physicists outside the experimental collaborations to question the reasons behind the initial discrepancies and their subsequent alignment\footnote {\textit{Whether this approach reinforces bias in data or data analyses against possible new physics, or whether no bias exists at all, is difficult to assess from our perspective. Our choice stands with LIV, so we prefer to be skeptical.}}.

Taking into account all above, our search strategy should account for the fact that all experimentally acquired data is a mixture of LI or almost LI data and the data with high rapidity, where LIV effects could, in principle, be detected. To enhance the likelihood of detecting Lorentz invariance violation (LIV), an effective approach would be to categorize the data by rapidity and sidereal time. This would enable event selection based on transferred momentum and spatial orientation. Smaller bins would give a better description, but on the other hand, higher rapidity cases, where the LIV effect is most pronounced, are naturally smaller in number. So, there might be a practical bin size limit we can afford. It would be easier to analyze pure time-like violation, since anisotropy is not present and, therefore, there is no necessity to divide the data by sidereal time. Therefore, in a time-like LIV scenario analysis, the number of events per bin will be greater than in a space-like LIV scenario analysis, although the total number of events remains the same.
In each bin, the cross-section distribution (near $M_Z$ region) as a function of invariant mass yields slightly different shapes, sizes, and peak locations , which enables identification of the LIV parameters ($\delta_{LIV}$, $n_{\mu}$).

\section{Conclusion}

{\normalsize In this work, we have explored the potential for detecting
Lorentz Invariance Violation (LIV) in the weak sector through modifications in the dispersion relation of the neutral Z boson. Our analysis shows that LIV can affect the decay rate (\ref{Decay L}) and propagator (\ref{pro2a}), consequently impacting the cross-section of Drell-Yan processes (\ref{D cross-section}) observed at proton-proton colliders (LHC, Tevatron). We identified that the effect is most pronounced near the resonance region at higher rapidities, where it alters both the resonance mass and value of the decay rate. }

We discussed three distinct cases of LIV: time-like, space-like, and light-like, each exhibiting unique signatures in experimental data. Our findings indicate that time-like violation amplifies with the rapidity of the Z boson, while space-like violation not only amplifies with rapidity but also introduces variations dependent on the experiment's orientation, leading to potential daily modulations due to the Earth's rotation. Light-like violation presents a hybrid case with more complex behavior influenced by both rapidity and spatial orientation.

To detect LIV, we propose dividing the experimental data by rapidity and sidereal time, thereby selecting events based on transferred momentum and spatial orientation. This approach enhances the likelihood of identifying
LIV effects, which may otherwise be lost in averaging procedures of traditional analysis. For pure time-like violation, our proposal offers a simplified scenario without anisotropy, eliminating the need for sidereal time division and yielding more robust statistics for each rapidity bin.

{\normalsize Our estimates performed for time-like LIV, suggest that a
perceived deviation $\Delta M_{Z}\approx-2.5$ $MeV$ could be observed for
$\delta_{LIV}=-10^{-8}$, corresponding value may remain noticeable even for
$\delta_{LIV}=-10^{-9},$ especially on the future upgraded LHC. This indicates that LIV effects of this magnitude are within reach of current experimental capabilities at the LHC, providing a promising avenue for future investigations into the fundamental nature of Lorentz invariance and its potential violations. }

\section{Acknowledgments}

{\normalsize This research was supported by a grant from the Shota Rustaveli National Science Foundation (SRNSF) under project number STEM-22-2604. We would like to express sincere gratitude to Professor John (Juansher) Chkareuli for his invaluable spiritual guidance and for instilling in us a passion for LIV from the very beginning of our careers.  We also want to extend our appreciation to Uta Klein, Enrico Lunghi, and Nathaniel Sherrill for their interesting discussions and shared interest for Lorentz Invariance Violation.}

\end{document}